\documentclass[12pt]{article}
\usepackage{latexsym}
\usepackage{amsmath}
\usepackage{amssymb}

\begin{document}
\thispagestyle{empty}

\centerline{\Large{\bf On Estimating the State of a Finite Level Quantum System}}
\centerline{\bf by}
\centerline{\bf K. R. Parthasarathy}
\centerline{\bf Indian Statistical Institute, Delhi Centre,}
\centerline{\bf 7, S. J. S. Sansanwal Marg, }
\centerline{\bf New Delhi - 110 016, India.}
\centerline{\bf e-mail : krp@isid.ac.in}
\vskip40pt

\vskip 0.5in
\noindent {\bf Summary :} We revisit the problem of mutually unbiased measurements in the context of estimating the unknown state of a $d$-level quantum system, first studied by W. K. Wootters and B. D. fields[7] in 1989 and later investigated by S. Bandyopadhyay et al [3] in 2001 and A. O. Pittenger and M. H. Rubin [6] in 2003. Our approach is based directly on the Weyl operators in the $L^2$-space over a finite field when $d=p^r$ is the power of a prime. When $d$ is not a prime power we sacrifice a bit of optimality and construct a recovery operator for reconstructing the unknown state from the probabilities of elementary events in different measurements.
\vskip 0.5in
\noindent AMS 2000 Mathematics Subject Classification 47L90, 47N50, 81P68 (?)

\section{Introduction}

This is almost an expository account of a well-known problem of quantum probability and statistics arising in the context of quantum information theory. There is a $d$-level quantum system whose pure states are described by unit vectors in a $d$-dimensional complex Hilbert space ${\cal H}$ equipped with the scalar product $\langle \varphi|\psi \rangle$ between elements $\varphi, \psi$ in ${\cal H}.$ This scalar product is linear  in the variable $\psi$ and antilinear in the variable $\varphi.$ Throughout this exposition we assume that $d$ is finite. Denote by ${\cal B}({\cal H})$ the $\star$-algebra of all operators on ${\cal H}.$ The complex $d^2$-dimensional vector space ${\cal B}({\cal H})$ will also be viewed as a Hilbert space with the scalar product $\langle X | Y \rangle = {\rm Tr} X^{\dagger} Y$ where $X^{\dagger}$ denotes the adjoint of the operator $X.$ Denote by ${\cal S}({\cal H}) \subset {\cal B}({\cal H})$ the compact convex set of all nonnegative (definite) operators of unit trace. Any element $\rho$ in ${\cal S}({\cal H})$ is called a {\it state} of the system. The extreme points of ${\cal S}({\cal H})$ are precisely one dimensional orthogonal projections. They are called {\it pure states.} In the Dirac notation any pure state can be expressed as $|\psi><\psi|$ where $\psi$ is a unit vector in ${\cal H}.$ Denote by ${\cal P}({\cal H})$ the set of all orthogonal projection operators (or, simply, projections) on ${\cal H}.$ Any element $P$ in ${\cal P}({\cal H})$ is called an {\it event} concerning the system and the quantity ${\rm Tr}\rho P$ is interpreted as the probability of the event $P$ in the state $\rho.$ In the context of quantum information theory the state of a quantum system can be utilized as an information resource. If the system is in an unknown state $\rho$ it is important to estimate $\rho$ from ``independent repeated measurements''. If we choose and fix an orthonormal basis $\{e_0, e_1, \ldots, e_{d-1}  \}$ in ${\cal H}$ then $\rho$ is described in this basis by a nonnegative definite matrix $((\rho_{ij}))$ where $\rho_{ij} = \langle e_i | \rho | e_j \rangle.$

Thus determination of $\rho$ involves the determination of $d^2-1$ real parameters, namely, $\rho_{ii}, i = 1, 2, \ldots, d-1,$ ${\rm Re}\,\rho_{ij},$ ${\rm Im} \rho_{ij},$ $0 \leq i < j \leq d-1.$ (Note that $\rho_{00} = 1 - \sum\limits_{i=1}^{d-1} \rho_{ii}$ and $\rho_{ij} = \bar{\rho}_{ji}.$)

By an {\it elementary measurement} ${\cal M} = \{ P_0, P_1, \ldots, P_{d-1} \}$ we mean a family of $d$ mutually orthogonal one dimensional projection operators $P_j, j=0,1,2,\ldots,d-1$ so that $\sum\limits_{0}^{d-1} P_j = I,$ the identity operator. If the measurement ${\cal M}$ is performed when the state of the system is $\rho,$ the result of such a measurement is one of the classical outcomes $j \in \{0,1,2, \ldots, d-1\}$ with probability ${\rm Tr} \rho P_j=p_j$ for each $j.$ Independent repeated trials of the measurement in the same state $\rho$ yield frequencies $f_j$ for each elementary outcome $j$ and $f_j$ can be viewed as an estimate of $p_j$ for each $j.$ Thus an elementary measurement covers at most $d-1$ degrees of freedom concerning $\rho$ in view of the relation $\sum\limits_{j=0}^{d-1} p_j = 1.$ In order to estimate $\rho$  it is therefore necessary to examine the frequencies of elementary outcomes in at least $d+1$ elementary measurements ${\cal M}_j, \,0 \leq j \leq d  $ where no two of the measurements ${\cal M}_i$ and ${\cal M}_j$  have any ``overlap of information''. Such an attempt is likely to cover all $(d+1)(d-1)= d^2-1$ degrees of freedom involved in reconstructing or estimating the unknown $\rho.$ To bring clarity to the notion of ``nonoverlap of information'' in a pair of elementary measurements it is useful to look at the $\star$-abelian algebra

$${\cal A} ({\cal M}) = \left \{\left . \sum_{j=0}^{d-1} a_j P_j \right | a_j \in {\Bbb C}, j = 0,1, \ldots, d-1    \right \}.$$
Any element $X = \sum\limits_{j=0}^{d-1} x_j P_j$ in ${\cal A} ({\cal M})$ can be looked upon as a complex-valued observable where $P_j$ is interpreted as the event that ``$X$ assumes the value $x_j$''. Of course, this is justified if all the $x_j$'s are distinct scalars. If $x$ is any scalar then the event that $X$ assumes the value $x$ is the projection $\sum\limits_{j:x_j=x} P_j.$ Thus the subalgebra ${\Bbb C }I \subset {\cal A} ({\cal M})$ consists precisely of constant-valued observables. Such an interpretation motivates the following formal definition.
\vskip10pt
\noindent{\bf Definition 1.1} Two elementary measurements ${\cal M} = \{ P_0, P_1, \ldots, P_{d-1} \},$ \\
${\cal M}^{\prime}= \{ Q_0, Q_1, \ldots, Q_{d-1} \} $ are said to be {\it weakly mutually unbiased} (WMUB) if
$${\cal A} ({\cal M}) \cap {\cal A} ({\cal M}^{\prime}) = {\Bbb C} I,$$
and {\it strongly mutually unbiased} (SMUB) if, in the Hilbert space ${\cal B}({\cal H}),$ the subspaces ${\cal A}({\cal M})\ominus {\Bbb C} I$ and  ${\cal A}({\cal M}^{\prime})\ominus {\Bbb C} I$ are mutually orthogonal. (Here, for two subspaces $S_1 \subset S_2 \subset {\cal B}({\cal H}), S_2 \ominus S_1$ denotes the orthogonal complement of $S_1$ in $S_2$). 

Clearly SMUB implies WMUB. We shall now describe these two properties in terms of the quantities ${\rm Tr} P_i Q_j.$
\vskip10pt
\noindent {\bf Proposition 1.2} Two elementary measurements ${\cal M} = \{ P_0, P_1, \ldots, P_{d-1} \},$ ${\cal M}^{\prime} = \{Q_0, Q_1, \ldots, Q_{d-1} \}$ are SMUB if and only if
\begin{equation}
{\rm Tr} P_i Q_j = d^{-1} \,\,\mbox{for all}\,\, i, j, \in \{ 0,1,2,\ldots, d-1 \}. \label{a1}
\end{equation}
\vskip10pt
\noindent{\bf Proof:} Note that the subspaces ${\cal A}({\cal M}) \ominus {\Bbb C} I$ and  ${\cal A}({\cal M}^{\prime}) \ominus {\Bbb C} I$ are respectively spanned by the subsets $\{P_j - d^{-1} I, \,0\leq j \leq d-1   \}$ and $\{Q_j - d^{-1}I, 0 \leq j \leq d-1 \}.$ Thus the orthogonality of these two subspaces is equivalent to the condition
$$0 = \langle \left . P_i - d^{-1} I \right | Q_j - d^{-1} I \rangle =\, {\rm Tr} (P_i -d^{-1} I) (Q_j - d^{-1}I)=({\rm Tr} P_i Q_j)-d^{-1} $$
for all $i,j$ in $\{0,1,2,\ldots, d-1\}.$ $\Box$
\vskip10pt
\noindent{\bf Proposition 1.3} Let ${\cal M} = \{P_0, P_1, \ldots, P_{d-1} \},$ ${\cal M}^{\prime} = \{ Q_0, Q_1, \ldots, Q_{d-1}\}$ be two elementary measurements. Suppose
$$L = \left [{\rm Tr}(P_i - P_0) (Q_j - Q_0) \right ], \,\, i,j \in \{ 1,2,\ldots,d-1\}$$
and $J_{d-1}$ is the $(d-1)\times (d-1)$ matrix all the entries of which are unity. Then ${\cal M}$ and ${\cal M}^{\prime}$ are WMUB if and only if
\begin{equation}
\det \left (I_{d-1} + J_{d-1} + d^{-1} L J_{d-1} L^{\dagger} - LL^{\dagger} \right ) > 0. \label{a2}
\end{equation}
\vskip10pt
\noindent{\bf Proof:} Let $X \in {\cal A} ({\cal M}) \cap {\cal A} ({\cal M}^{\prime}).$ Then there exist scalars $a_i,b_j,$ $i,j \in \{1,2,\ldots,d-1 \}$ such that
\begin{eqnarray*}
X &=& d^{-1} ({\rm Tr}X) I + \sum_{i=1}^{d-1} a_i (P_i - P_0)\\
&=& d^{-1} ({\rm Tr}X) I + \sum_{j=1}^{d-1} b_j (Q_j - Q_0).
\end{eqnarray*} 
Thus ${\cal M}$ and ${\cal M}^{\prime}$ are WMUB if and only if the set $\{P_1-P_0, P_2-P_0, \ldots, P_d-P_0, Q_1-Q_0, Q_2- Q_0, \ldots, Q_d - Q_0 \}$ of $2(d-1)$ elements in the Hilbert space ${\cal B}({\cal H})$ is linearly independent. This, in turn, is equivalent to the strict positive definiteness of the partitioned matrix
$$\left [\begin{array}{c|c}[{\rm Tr} (P_i - P_0)(P_j-P_0)] & [{\rm Tr} (P_i-P_0)(Q_j-Q_0)] \\ \hline [{\rm Tr}(Q_i-Q_0)(P_j-P_0)] & [{\rm Tr}(Q_i-Q_0)(Q_j-Q_0)]  \end{array}  \right ], i,j \in \{1,2,\ldots, d-1\}$$
of order $2(d-1).$ We have
$${\rm Tr} (P_i-P_0)(P_j-P_0)={\rm Tr} (Q_i-Q_0)(Q_j-Q_0) = \left \{\begin{array}{ll}2 &\, \mbox{if}\,\,\, i=j,\\ 1 & \,\mbox{if}\,\,\, i \neq j.  \end{array}  \right . $$
Thus, ${\cal M}$ and ${\cal M}^{\prime}$ are WMUB if and only if
$$\left [\begin{array}{c|c}I_{d-1} + J_{d-1} & L \\ \hline L^{\dagger} & I_{d-1} + J_{d-1}  \end{array}   \right ]$$
has a strictly positive determinant. Left multiplication of this matrix by the matrix
$$\left [\begin{array}{c|c}I_{d-1} & - L (I_{d-1} + J_{d-1})^{-1} \\ \hline 0 & I_{d-1}   \end{array} \right ]$$
with unit determinant yields the equivalent condition
\begin{equation}
\det \left (I_{d-1} + J_{d-1} - L (I_{d-1} + J_{d-1})^{-1} L^{\dagger} \right ) > 0. \label{a3}
\end{equation}
Since
$$(I_{d-1} + J_{d-1})^{-1} = I_{d-1} - d^{-1} J_{d-1},$$
condition (\ref{a3}) reduces to condition (\ref{a2}). $\Box$
\vskip10pt
\noindent{\bf Corollary 1.4} If the matrix $L$ of Proposition 1.3 satisfies the inequality $\|L\| < 1$(where  $\| . \|$ is the standard operator norm in the $\star$-algebra ${\cal B}({\Bbb C}^{d-1})$ then ${\cal M},$ ${\cal M}^{\prime}$ are WMUB. Furthermore ${\cal M,}$ ${\cal M}^{\prime}$ are SMUB if and only if $L=0.$
\vskip10pt
\noindent{\bf Proof:} Immediate. $\Box$

In the context of minimizing the number of elementary measurements required for estimating the state $\rho$ of a quantum system Proposition 1.2 emphasizes the importance of the search for $d+1$ elementary measurements which are pairwise SMUB. When $d$ is a prime power $p^r$ the existence of such a family of SMUB measurements was proved by Wootters and Fields [7]. Alternative proofs of this result were given by S. Bandyopadhyay et al in [3] and Pittenger and Rubin in [6]. In this paper we shall present a proof of the same result by using the commutation relations of Weyl operators in the $L^2$ space of the finite field ${\Bbb F}_{P^{r}}.$ When $d = p_1^{m_{1}}  p_2^{m_{2}} \ldots  p_n^{m_{n}}$ with $p_i$'s being prime we shall use the Weyl commutation relations in the $L^2$ space of the additive abelian group $\otimes_{i=1}^n {\Bbb F}_{{p_{i}}^{m_{i}}}$ and study the problem of estimating the unknown state of a $d$-level system. This leads to an interesting reconstruction formula for a state $\rho$ in terms of probabilities of $d^2-1$ events arising from $\prod_{i=1}^n (p_i^{m_{i}} + 1)$ elementary measurements. However, one would like to express $\rho$ in terms of the probabilities of elementary outcomes in  $\left (\prod_{i=1}^n p_i ^{m_{i}} + 1  \right )$ elementary measurements.

\section{The case $d = p^r$} 
\setcounter{equation}{0}

Let $\dim {\cal H} = d = p^r$ be a prime power. For any prime power $q$ denote by ${\Bbb F}_q$ the unique (upto a field isomorphism) finite field of cardinality $q.$ Choose and fix any nontrivial character $\chi$ of the additive group ${\Bbb F}_d$ and put
\begin{equation}
\langle x, y \rangle = \chi (xy), \,\, x, y \in {\Bbb F}_d. \label{b1}
\end{equation}
One can, for example, look upon ${\Bbb F}_d$ as an $r$-dimensional vector space over ${\Bbb F}_p,$ express any element $x$ in ${\Bbb F}_d$ as an ordered $r$-tuple: $x = (s_1, s_2, \ldots, s_r)$ where $0 \leq s_i \leq p-1$ for each $i$ and put
\begin{equation}
\chi (x) = \exp \frac{2 \pi i}{p} s_1. \label{b2}
\end{equation}
Then we have $|\langle x, y \rangle | = 1,$ $\langle x, y \rangle = \langle y, x \rangle,$ $\langle x, y_1+y_2 \rangle = \langle x, y_1 \rangle \langle x, y_2 \rangle$ and $x=0$ if $\langle x, y \rangle = 1$ for all $y$ in ${\Bbb F}_d.$ In other words, $\langle .,.\rangle$ is a nondegenerate symmetric bicharacter for ${\Bbb F}_d.$ Identify the Hilbert space ${\cal H}$ with $L^2 ({\Bbb F}_d),$ using the counting measure in ${\Bbb F}_d,$ and put
$$| x > = 1_{\{x\}}, x \in {\Bbb F}_d$$
where $1_{\{x\}}$ is the indicator function of the singleton subset $\{x\}$ in ${\Bbb F}_d.$ Then $\{|x>, x \in {\Bbb F}_d \}$ is an orthonormal basis for ${\cal H}$ labelled by the elements of ${\Bbb F}_d.$ Now, consider the unique unitary operators $U_a,$ $U_b$ in ${\cal H}$ determined by the relations
\begin{eqnarray*}
U_a | x > &=& | a + x >, \\
V_b | x > &=& <b,x> | x > \quad \mbox{for all}\,\, x \in {\Bbb F}_d.
\end{eqnarray*}
Then we have
\begin{eqnarray}
U_a U_b &=& U_{a+b}, V_a V_b = V_{a+b}, \label{b3}\\
V_b U_a &=& \langle a, b \rangle U_a V_b. \label{b4}
\end{eqnarray}
Elementary algebra shows that
\begin{equation}
{\rm Tr}\,(U_{a_{1}} V_{b_{1}})^{\dagger} U_{a_{2}} V_{b_{2}} = d \delta_{a_{1}, a_{2}} \delta_{b_{1}, b_{2}} \label{b5}
\end{equation}
for all $a_1, a_2, b_1, b_2$ in ${\Bbb F}_d.$ In particular, the family $\{U_a V_b, a,b \in {\Bbb F}_d  \}$ of $d^2$ unitary operators constitute an orthogonal basis for the Hilbert space ${\cal B}({\cal H}).$ This is an example of a unitary error basis in the theory of error correcting quantum codes [4]. Notice also the fact that $\{U_a\}$ and $\{V_b\}$ are like the position and momentum representations obeying the Weyl commutation relations in classical quantum mechanics. In view of this property we call any operator of the form $\lambda U_a V_b,$ $|\lambda|=1,$ $a,b \in {\Bbb F}_d$ a {\it Weyl operator.} We say that (\ref{b3}) and  (\ref{b4}) constitute the {\it Weyl commutation relations.} The usefulness of such an error basis of Weyl operators in the study of quantum codes has been explored in [1], [2],[5]. We shall slightly modify the error basis $\{U_a V_b\}$ by multiplying each element $U_a V_b$ by an appropriate phase factor. Once again viewing ${\Bbb F}_d$ as an $r$-dimensional vector space over ${\Bbb F}_p,$ expressing any $x \in {\Bbb F}_d$ as an ordered $r$-tuple $x = (s_1, s_2, \ldots, s_r)$ with $0 \leq s_i \leq p-1$ for each $i$ and considering the basis elements $e_i = (0,0,\ldots,0,1,0,\ldots,0)$ of the field ${\Bbb F}_d$ with $1$ in the $i$-th position and $0$ elsewhere we write $x = s_1 e_1 + s_2 e_2 + \cdots + s_r e_r$ and define
\begin{equation}
\alpha (a,x) = \chi \left (a \left \{\sum_{i < j} s_i s_j e_i e_j + \sum_j \frac{s_j (s_j-1)}{2} e_j^2 \right \}   \right ), a_j x \in {\Bbb F}_d \label{b6}
\end{equation}
where $\chi$ is the character chosen and fixed at the beginning of this section.

Now put $\bar{{\Bbb F}}_d = {\Bbb F}_d \cup \{\infty\}  $ and write
\begin{equation}
W(a,x) = \left \{\begin{array}{lcl} \alpha (a, x) U_x V_{ax} &\mbox{if}& a \in {\Bbb F}_d, x \in {\Bbb F}_d, \\ V_x & \mbox{if} & a = \infty.      \end{array}  \right . \label{b7}
\end{equation}
Then we have the following proposition.
\vskip10pt
\noindent{\bf Proposition 2.1} The family $ \left \{I, W(a,x), a \in \bar{{\Bbb F}}_d, x \in {\Bbb F}_d \setminus \{0\}  \right \}$ is an orthogonal basis of unitary operators for the operator Hilbert space ${\cal B}({\cal H})$ satisfying the relations
\begin{equation}
W (a, x) W(a, y) = W(a, x+y) \,\,\mbox{for all}\,\, a \in \bar{{\Bbb F}}_d, x \in {\Bbb F}_d. \label{b8}
\end{equation}
\vskip10pt
\noindent {\bf Proof :} The first part is immediate from the fact that the family of operators under consideration differs from the family $\{U_x V_y,  x, y \in {\Bbb F}_d\}$ only by a scalar factor of modulus unity in each element. If $a \in {\Bbb F}_d,$ $x = \sum s_i e_i,$ $y = \sum t_i e_i$ we have from (\ref{b3}) (\ref{b4})
\begin{eqnarray*}
\lefteqn{W (a,x) W (a,y)                } \\
&=& \alpha (a,x) \alpha (a,y) \langle ax, y \rangle U_{x + y} V_{a(x+y)}  \\
&=& \alpha (a,x) \alpha (a,y) \overline{\alpha (a, x+ y)} \langle ax, y \rangle W (a, x+y)
\end{eqnarray*}
where the coefficient of $W(a, x+y)$ is of the form $\chi (a z)$ with
\begin{eqnarray*}
 z &=& \sum_{i < j} s_i s_j e_i e_j + \sum_j \frac{s_j (s_j-1)}{2} e_j^2 + \sum_{i< j} t_i t_j e_i e_j + \sum_j \frac{t_j (t_j-1)}{2} e_j^2  \\
&&- \sum_{i < j} (s_i + t_i)(s_j+t_j) e_i e_j -\sum_j \frac{(s_j + t_j)(s_j + t_j -1)}{2}  e_j^2 + \sum_{i,j} s_i t_j e_i e_j   \\
&=& 0.
\end{eqnarray*}
This proves (\ref{b8}) when $a \in {\Bbb F}_d.$ When $a = \infty,$ (\ref{b8}) is a part of (\ref{b3}). $\Box$
\vskip10pt
\noindent{\bf Theorem 2.2} There exists a family of one dimensional orthogonal projection operators $\{P (a,x), a \in \bar{{\Bbb F}}_d, x \in {\Bbb F}_d\}$ satisfying the following :

\begin{itemize}
\item [(i)] $W(a,x) = \sum\limits_{y \in {\Bbb F}_d} \langle x, y \rangle P(a,y)   $
\item[(ii)] $P (a,y) = d^{-1} \sum\limits_{x \in {\Bbb F}_d} \overline{\langle x, y \rangle  } W(a, x),$
\item [(iii)] $P(a,x) P(a,y) = \delta_{x,y} P(a,x),  $
\item[(iv)] $\sum_{x \in {\Bbb F}_d } P(a, x) = I,$
\item[(v)] ${\rm Tr}\,P (a,x) P(b,y) = d^{-1}$ for all $a \neq b;$ $a, b \in \bar{{\Bbb F}}_d;$ $x, y \in {\Bbb F}_d.$
\end{itemize}

\vskip10pt
\noindent{\bf Proof :} By Proposition 2.1 the correspondence $x \rightarrow W (a,x)$ is a unitary representation of the additive abelian group ${\Bbb F}_d$ and $\{\langle ., y \rangle, y \in {\Bbb F}_d \}$ is the set of all its characters. Thus the decomposition of $\{ W(a,.)\}$ into its irreducible components yields a spectral measure $P(a,.)$ on ${\Bbb F}_d$ satisfying (i), (iii) and (iv). Substituting from (i) the expression for $W(a,x)$ in the right hand side of (ii) and using the orthogonality relations for characters we get (ii). Taking trace on both the sides of (ii) and observing that $W (a,0) = I$ and ${\rm Tr}\,W (a,x) = 0$ for $x \neq 0$ we get ${\rm Tr}\, P (a,y)=1.$ Thus each $P(a,y)$ is a one dimensional projection. Substituting for $P(a, x)$ and $P(a,y)$ from (ii) in the left hand side of (v) we have from (\ref{b7}), (\ref{b3}) and (\ref{b4})
\begin{eqnarray*}
\lefteqn{ {\rm Tr}\, P(a,x) P(b,y)                     }  \\
&=& d^{-2}  \sum\limits_{z_1, z_2 \in {\Bbb F}_d} \langle x, z_1 \rangle \langle y, z_2 \rangle \,\,{\rm Tr}\,\, W (a,z_1) W (b, z_2)\\
&=& d^{-2}  \sum\limits_{z_1, z_2 \in {\Bbb F}_d} \langle x, z_1 \rangle \langle y, z_2 \rangle \alpha (a, z_1) \alpha (b, z_2) \langle az_1, z_2 \rangle \,{\rm Tr}\, U_{z_{1} + z_{2}} V_{az_{1} + bz_{2}}.
\end{eqnarray*}
Now observe that the $(z_1, z_2)$-th term of the sum on the right hand side is nonzero only if $z_1 + z_2 = 0,$ $az_1 + bz_2 = 0.$ If $a \neq b$ this is possible only if $z_1 = z_2 = 0.$ This proves (v). $\Box$
\vskip10pt
\noindent{\bf Corollary 2.3} Let ${\cal M}_a = \{P(a,x), x \in {\Bbb F}_d\}.$ Then $\{{\cal M}_a, a \in \bar{{\Bbb F}}_d \}$ is a set of $(d+1)$ elementary measurements which are pairwise SMUB.
\vskip10pt
\noindent {\bf Proof:} Immediate from Propostion 1.2. $\Box$

Our next result yields a recovery formula for any state $\rho$ from the probability distributions $\{{\rm Tr}\, \rho P(a,x), x \in {\Bbb F}_d\}$ on ${\Bbb F}_d$ arising from the measurements $\{{\cal M}_a,  a \in \bar{{\Bbb F}}_d\}.$
\vskip10pt
\noindent{\bf Theorem 2.4} Let $\{P(a,x), a \in \bar{{\Bbb F}}_d, x \in {\Bbb F}_d  \}$ be the projections in Theorem 2.2. Then, for any state $\rho$ on $L^2 ({\Bbb F}_d)$ the following holds:
\begin{itemize}
\item[(i)] $\rho = \sum\limits_{a \in \bar{{\Bbb F}}_d} \sum\limits_{z \in {\Bbb F}_d} \{{\rm Tr}\, \rho P(a,z) - \frac{1}{d+1}  \} P(a,z)    $
\item[(ii)] $\rho = \sum\limits_{x, y \in {\Bbb F}_d \atop  {a \in \bar{{\Bbb F}}_d}} \overline{\langle x, y \rangle} \{{\rm Tr}\, \rho P(a,y) \} W(a,x)  $
\end{itemize}
\vskip10pt
\noindent {\bf Proof:} From the first part of Proposition 2.1, it follows that $\rho$ admits the expansion
$$\rho = d^{-1} \left \{I +  \sum\limits_{a \in \bar{{\Bbb F}}_d \atop {x \in {\Bbb F}_d \setminus \{0\}}}  \left [{\rm Tr}\,\rho W (a,x)^{\dagger} \right ] W (a,x) \right \}$$
in terms of the orthogonal basis arising from the Weyl operators. Now substitute in the right hand side the expressions for $W(a,x)$ in (i) of Theorem 2.2 and use the orthogonality relations for characters:
$$\sum_{x \in {\Bbb F}_d} \overline{\langle x,y \rangle} \langle x, z \rangle  = d \delta_{y,z}$$
Then we obtain the identity (i) of the theorem. If we substitute for $P(a,z)$ from the identity (ii) of Theorem 2.2 we obtain the second identity of the theorem. $\Box$
\vskip10pt
\noindent {\bf Remark:} If we make repeated independent measurements ${\cal M}_a,$ obtain the frequencies for the different events $P(a,z)$ and substitute those frequencies for the different probabilities ${\rm Tr}\, \rho P (a,z)$ in the unknown state $\rho$ we will get an unbiased and asymptotically consistent estimate $\hat{\rho}$ of $\rho$ but $\hat{\rho}$ may not be a positive operator. One may replace $\hat{\rho}$ by the normalised version of the positive part or the modulus of $\hat{\rho}$ at the cost of losing unbiasedness. This also increases the computational cost.

\section{Estimation of states in the general case}
\setcounter{equation}{0}

Let  $d = p_1^{m_{1}} p_2^{m_{2}} \ldots p_n^{m_{n}}$ be the decomposition of $d$ into its prime factors $p_1 < p_2 < \cdots < p_n.$ Write $d_j = p_j^{m_{j}}.$ We may identify the $d$-dimensional Hilbert space ${\cal H}$ with ${\cal H}_1 \otimes {\cal H}_2 \otimes \cdots \otimes {\cal H}_n$ where ${\cal H}_j = L^2 ({\Bbb F}_{d_{j}}),$ ${\Bbb F}_{d_{j}}$ being the finite field of cardinality $d_j.$ Following the definition in (\ref{b7}) construct the unitary operators $W^{(j)} (a_j, x_j)$ when $d = d_j, \,j=1,2,\ldots, n$ and using Theorem 2.2, the corresponding projections $P^{(j)} (a_j, x_j),$ where $a_j \in {\Bbb F}_{d_{j}},$ $x_j \in {\Bbb F}_{d_{j}}.$ We now adopt the following convention: for any operator $X$ in $L^2 ({\Bbb F}_{d_{j}}) = {\cal H}_j$ denote by the same symbol $X$ the operator in ${\cal H}$ defined by $X = X_1 \otimes X_2 \otimes \cdots \otimes X_n$ where $X_i$ is the identity operator in ${\cal H}_i$ when $i \neq j$ and $X_j = X.$ The operator $X$ thus defined in ${\cal H} = {\cal H}_1 \otimes {\cal H}_2 \otimes \cdots \otimes {\cal H}_n$ is called the {\it ampliation} of $X$ in ${\cal H}_j$ to ${\cal H}.$ Since ${\cal B}({\cal H})$ can be identified with ${\cal B}({\cal H}_1) \otimes {\cal B}({\cal H}_2) \otimes \cdots \otimes {\cal B}({\cal H}_n)$ as Hilbert spaces as well as $\star$-algebras it follows from Proposition 2.1 that the family
\begin{eqnarray}
{\cal F} &=& \left \{ I, W^{(i_{1})} (a_{i_{1}}, x_{i_{1}}) W^{(i_{2})} (a_{i_{2}}, x_{i_{2}}) \cdots  W^{(i_{r})} (a_{i_{r}}, x_{i_{r}}),   \right . \nonumber \\
&&a_{i_{j}} \in  \bar{{\Bbb F}}_{d_{j}}, x_{i_{j}} \in {\Bbb F}_{d_{j}} \setminus \{0\}, j = 1,2,\ldots, r,     \nonumber \\
&& \left . 1 \leq i_1 < i_2 < \cdots < i_r \leq n, r=1,2,\ldots, n    \right \} \label{c1}
\end{eqnarray}
of unitary operators in ${\cal H}$ constitute an orthogonal basis for the operator Hilbert spaces ${\cal B}({\cal H}).$ Note that the cardinality of ${\cal F}$ is, indeed, equal to
\begin{eqnarray*}
\lefteqn{1 + \sum_{r=1}^n \sum_{1 \leq i_1 < i_2 < \cdots < i_r \leq n } (d_{i_{1}}^2 -1)(d_{i_{2}}^2 -1) \ldots (d_{i_{r}}^2 -1)       } \\
&=& (1 + d_1^2 -1)(1+ d_2^2-1) \ldots (1 + d_n^2 -1) \\
&=&d_1^2 d_2^2 \ldots d_n^2 \\
&=& d^2,
\end{eqnarray*}
the dimension of ${\cal B}({\cal H}).$ For any subset $J = \{i_1, i_2, \ldots, i_r \} \subset \{1,2,\ldots,n\}$ where $1 \leq i_1 < i_2 < \cdots < i_r \leq n,$ define
\begin{eqnarray*}
d(J) &=& d_{i_{1}} d_{i_{2}} \ldots d_{i_{r}}\\
d^{\prime}(J) &=& (d_{i_{1}}+1)(d_{i_{2}}+1) \cdots (d_{i_{r}}+1), 
\end{eqnarray*}
and for any state $\rho$ in ${\cal H},$ put
\begin{eqnarray}
S_{\rho}(J) &=& \sum_{a_{i_{j}} \in \bar{{\Bbb F}}_{d_{i_{j}}},  \atop {y_{i_{j}} \in {\Bbb F}_{d_{i_{j}}}}  \forall j  } \left \{{\rm Tr}\, \rho P^{(i_{1})} (a_{i_{1}}, y_{i_{1}}) P^{(i_{2})}  (a_{i_{2}}, y_{i_{2}}) \ldots P^{(i_{r})}  (a_{i_{r}}, y_{i_{r}})   \right \} \nonumber \\
&& P^{(i_{1})}  (a_{i_{1}}, y_{i_{1}}) P^{(i_{2})}  (a_{i_{2}}, y_{i_{2}}) \ldots P^{(i_{r})}  (a_{i_{r}}, y_{i_{r}}) \label{c2}
\end{eqnarray}
where $\{P^{(i)} (a_i,  y_i)\}$ are the one dimensional projections in ${\cal H}_i$ determined by the unitary representation $x_i \rightarrow W^{(i)} (a_i, x_i)$ of the additive group ${\Bbb F}_{d_{i}}$ according to Theorem 2.2 and ampliated to the product Hilbert space ${\cal H} = {\cal H}_1 \otimes {\cal H}_2 \otimes \cdots \otimes {\cal H}_n.$ Thus $S_{\rho} (J)$ is an operator in ${\cal H}$ determined by the probabilities ${\rm Tr}\, \rho P^{(i_{1})}  (a_{i_{1}} y_{i_{1}}) P^{(i_{2})}  (a_{i_{2}} y_{i_{2}}) \ldots P^{(i_{r})}  (a_{i_{r}} y_{i_{r}})$ and  the projections $P^{(i_1)} (a_{i_{1}}, y_{i_{1}}) P^{(i_{2})}  (a_{i_{2}}, y_{i_{2}}) \ldots  P^{(i_{r})}  (a_{i_{r}}, y_{i_{r}})$ of dimension $\Pi_{j \not\in \{i_1, i_2, \cdots, i_r\}} d_j$ with $a_i$'s varying in $\bar{{\Bbb F}}_{d_{i}}$ and $y_i$'s in ${\Bbb F}_{d_{i}}$ for any $i.$ With these notations and the convention $S_{\rho}(\phi)=I,$ we have the following theorem for the recovery of $\rho$ from the probabilities.
\vskip10pt
\noindent{\bf Theorem 3.1} Let $\rho$ be any state in ${\cal H}.$ Then
\begin{equation}
\rho = \sum_{J \subset \{1,2,\ldots,n \}} (-1)^{n-|J|} S_{\rho}(J) \label{c3}
\end{equation}
where $S_{\rho}(J)$ is given by (\ref{c2}) and $|J|$ is the cardinality of $J.$

\vskip10pt
\noindent{\bf Proof:} Since the family ${\cal F}$ of unitary operators in (\ref{c1}) is an orthogonal basis for ${\cal B}({\cal H})$ we can expand the state $\rho$ in this basis as
\begin{eqnarray}
\rho &=& (d_1 d_2 \ldots d_n)^{-1} \left \{ I + \sum_{r=1}^n \sum_{1 \leq i_1 < i_2 < \cdots < i_r \leq n} \sum_{a_{i_{j}} \in \bar{{\Bbb F}}_{d_{i_{j}}}, x_{i_{j}} \in {\Bbb F}_{d_{i_{j}}} \setminus \{0\} }    \right . \nonumber \\
&&\left . \left [{\rm Tr}\, \rho W^{(i_{1})} (a_{i_{1}}, x_{i_{1}})^{\dagger} \cdots W^{(i_{r})} (a_{i_{r}}, x_{i_{r}})^{\dagger}  \right ]W^{(i_{1})} (a_{i_{1}}, x_{i_{1}}) \cdots W^{(i_{r})} (a_{i_{r}}, x_{i_{r}}) \right \}.\label{c4}
\end{eqnarray}
From Theorem 2.2 we have for any fixed $i$
\begin{eqnarray*}
\lefteqn{\sum_{x_i \in {{\Bbb F}}_{d_{i}} \setminus \{0\}}  W^{(i)} (a_i, x_i)^{\dagger} \otimes W^{(i)} (a_i, x_i) }\\
&=& \sum_{y, z \in {{\Bbb F}}_{d_{i}} \atop{x_i \in {\Bbb F}_{d_{i}} \setminus \{0\} }} \overline{\langle x_i, y \rangle} \langle x_i, z \rangle P^{(i)} (a_i, y) \otimes P^{(i)} (a_i, z)                         \\
&=& d_i \sum_{y \in {{\Bbb F}}_{d_{i}}} P^{(i)} (a_i, y) \otimes P^{(i)} (a_i, y) - I^{(i)} \otimes I^{(i)},
\end{eqnarray*}
$I^{(i)}$ being the identity operator in ${\cal H}_i.$ Using this identity and elementary properties of relative trace, equation (\ref{c4}) can be written as
\begin{eqnarray*}
\rho &=& (d_1 d_2 \ldots d_n)^{-1}  \sum_{J} \sum_{K \subset J} (-1)^{|J|-|K|} d (K) d ^{\prime} (J \setminus K)\\
&\times& \sum_{a_{k_{i}} \in \bar{{\Bbb F}}_{d_{k_{i}}}, \atop{y_{k_{i}}} \in {\Bbb F}_{d_{k_{i}}} \forall i}\left \{{\rm Tr}\, \rho P^{(k_{1})} (a_{k_{1}}, y_{k_{1}})P^{(k_{2})} (a_{k_{2}}, y_{k_{2}}) \cdots P^{(k_{s})} (a_{k_{s}}, y_{k_{s}})     \right \} \\
&& \times P^{(k_{i})} (a_{k_{1}}, y_{k_{1}}) P^{(k_{2})} (a_{k_{2}}, y_{k_{2}}) \cdots P^{(k_{s})} (a_{k_{s}}, y_{k_{s}})
\end{eqnarray*}
where $J$ varies over all subsets $i_1 < i_2 < \cdots < i_r$ of $\{1,2,\ldots, n\}$ and $K$ varies over all subsets $k_1 < k_2 < \cdots < k_s$ of $J.$ Now using the definition in (\ref{c2}) we can express $\rho$ as
$$\rho = \sum_{K \subset \{1,2,\ldots, n\}} \alpha (K) S_{\rho}(K)$$
where
\begin{eqnarray*}
\alpha(K) &=& (d_1 d_2 \ldots d_n)^{-1} d(K) \sum_{L:L\cap K=\phi}(-1)^{|L|} d^{\prime}(L)\\
&=& (-1)^{n-|K|}. \hfill{\Box}
\end{eqnarray*}
\vskip10pt
\noindent{\bf Remark} From Theorem 3.1 it is clear that $\rho$ is recovered from the probabilities for the elementary events
$$P^{(1)} (a_1, x_1) P^{(2)} (a_2, x_2) \ldots P^{(n)} (a_n, x_n), \quad a_i \in \bar{{\Bbb F}}_{d_{i}}  x_i \in \bar{{\Bbb F}}_{d_{i}}.$$
In other words the determination of $\rho$ involves $(d_1+1)(d_2+1) \cdots (d_n+1)$ elementary measurements. As mentioned in the introduction one would like to determine $\rho$ by $d_1 d_2 \ldots d_n +1$ measurements.
\vskip20pt
\noindent{\bf Acknowledgement:} I wish to thank Professor S. Chaturvedi of the University of Hyderabad for bringing my attention to the central problem of this paper and the reference [7].
\vskip0.5in
\noindent {\bf References}
\vskip20pt
\begin{enumerate}
\item V. Arvind and K. R. Parthasarathy, {\it A family of quantum stabilizer codes based on the Weyl commutation relations over a finite field,} in A Tribute to C. S. Seshadri, Perspectives in Geometry and Representation Theory, Ed. V. Lakshmibai et al, Hindustan Book Agency, New Delhi (2003) 133-149.

\item V. Arvind, P. Kurur and K. R. Parthasarathy, {\it Nonstabilizer quantum codes from abelian subgroups of the error group,} quant-ph/0210097, to appear in Volume in honour of A. S. Holevo on his 60th birthday Ed. O. Hirota, 2004.

\item S. Bandyopadhyay, P. O. Boykin, V. Roychowdhury and F. Vatan, {\it A new proof for the existence of mutually unbiased bases}, arXiv:quant-ph/0103162 v3, 7 Sept.2001.

\item M. A. Nielsen and I. L. Chuang, {\it Quantum Computation and Quantum Information,} Cambridge University Press, 1999.

\item K. R. Parthasarathy, {\it Lectures on quantum computation, quantum error-correcting codes and information theory} (Notes by Amitava Bhattacharyya, TIFR, Mumbai, 2003).

\item A. O. Pittenger and M. H. Rubin, {\it Mutually unbiased bases, generalized spin matrices and separability,} arXiv:quant-ph/0308142 v1, 26 August 2003.

\item W. K. Wooters and B. D. Fields, {\it Optimal state-determination by mutually unbiased measurements,} Annals of Physics, 191 (1989) No.2, 363-381.

\end{enumerate}

\end{document}